\documentclass[draft]{agujournal2019}
\usepackage{rotating}
\usepackage{graphicx}

\setkeys{Gin}{draft=false}
\journalname{Geophysical Research Letters}

\begin{document}
\title{Flux erosion of magnetic clouds by reconnection with the Sun's open flux}

\authors{Sanchita Pal\affil{1}, Soumyaranjan Dash\affil{1}, and Dibyendu Nandy\affil{1,2}}
\affiliation{1}{Center of Excellence in Space Sciences India, Indian Institute of Science Education and Research Kolkata, Mohanpur 741246, West Bengal, India} 
\affiliation{2}{Department of Physical Sciences, Indian Institute of Science Education and Research Kolkata, Mohanpur 741246, West Bengal, India}

\correspondingauthor{Dibyendu Nandy}{dnandi@iiserkol.ac.in}
\begin{keypoints}

\item Flux of magnetic clouds is eroded due to magnetic reconnection with the solar open flux, impacting their geoeffectiveness in some cases.

\item A linear relationship is observed between eroded and open flux with larger average flux erosion during the stronger cycle 23 compared to 24.

\item We establish a solar cycle link to the evolution of magnetic cloud flux during their interplanetary propagation.

\end{keypoints}

\section*{Abstract}
Magnetic clouds (MCs) are flux-rope magnetic structures forming a subset of solar coronal mass ejections which have significant space weather impacts. The geoeffectiveness of MCs depends on their properties which evolve during their interplanetary passage. Based on an analysis of observations spanning two solar cycles we establish that MCs interacting with the ambient solar wind magnetic field (i.e., heliospheric open flux) lose a substantial amount of their initial magnetic flux via magnetic reconnection, which in some cases, reduce their geoeffectiveness. We find a linear correlation between the eroded flux of MCs and solar open flux which is consistent with the scenario that MC erosion is mediated via the local heliospheric magnetic field draping around an MC during its interplanetary propagation. The solar open flux is governed by the sunspot cycle. This work therefore uncovers a hitherto unknown pathway for solar cycle modulation of the properties of MCs.

\section*{Plain Language Summary}
Vast eruptions of magnetized plasma from the Sun known as coronal mass ejections (CMEs) are the main contributors to severe space weather conditions. The magnetic properties of CMEs -- which determine their geoeffectiveness -- evolve during their interplanetary journey. Therefore it is important to understand the physical processes that govern their evolution. Based on in situ satellite observations, and modeling here we provide evidence that the magnetic flux of CME associated magnetic clouds erode over their interplanetary journey due to reconnection with the ambient heliospheric open flux. We also demonstrate that this flux erosion impacts the geoeffectiveness of magnetic clouds in some cases. This mechanism establishes a novel connection between the solar cycle variation of the heliospheric open flux and space weather in the vicinity of solar system planets.\\

\section{Introduction}

\begin{figure}[htbp]
 \centering
 \includegraphics[width=\textwidth]{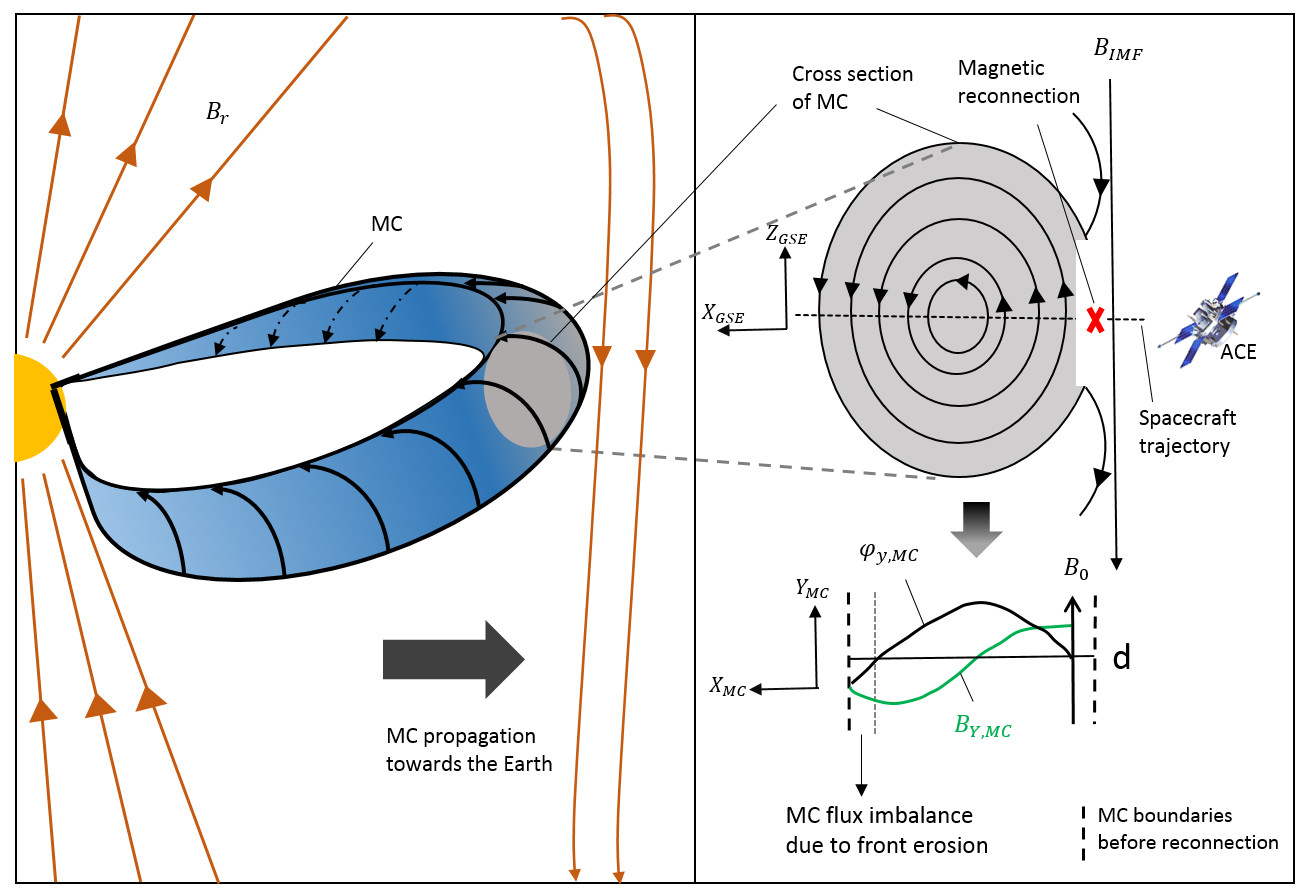}
 \caption{An idealized schematic of the ambient IMF draping about an MC propagating radially outward from the Sun towards the Earth (left), the MC's magnetic structure in a plane perpendicular to the ecliptic plane (right-top) and the expected variation in the MC's accumulated azimuthal flux due to its reconnection with the draped ambient IMF (right-bottom)}.
 \label{s1}
\end{figure}

Coronal mass ejections (CMEs) are magnetised plasma structures expelled sporadically from the Sun and travel toward the heliosphere as interplanetary coronal mass ejections (ICMEs). If an ICME has a smoothly rotating strong magnetic field and a low proton temperature \cite{1981JGR....86.8893B,1982JGR....87..613K} it is categorised as a magnetic cloud (MC). The twisted flux ropes of MCs \cite{1983NASCP.2280.731G,1988JGR....93.7217B,1990JGR....9511957L} carry a significant amount of solar magnetic flux and helicity \cite{2017ApJ...851..123P}. While propagating through the interplanetary medium, the presence of a relative motion between an MC and the ambient plasma would result in a draping of the ambient interplanetary magnetic field (IMF) over the MC \cite{1987GeoRL..14..355G}. Such draping has been directly observed (e.g., \citeA{crooker1985magnetic})
as well as modeled (e.g., \citeA{alksne1970magnetic}). If the magnetic field orientation of the MC is opposite to that of the draped IMF, the process of magnetic reconnection (henceforth reconnection) is initiated \cite{1988JGR....93.2519M}. Due to this reconnection a part of the MC's magnetic flux is eroded away and an asymmetry is generated in its azimuthal magnetic flux that accumulates along the spacecraft trajectory. This fact has been established by observations \cite{2006A&A...455..349D,2007SoPh..244..115D,2008AnGeo..26.3139M,2012JGRA..117.9101R,2015JGRA..120...43R} as well as by using global MHD simulations \cite{2003JGRA..108.1023S,angeo-28-1075-2010}. \citeA{2014JGRA..119...26L} showed that  erosion of an MC’s magnetic flux may reduce its geoeffectiveness by a significant amount. Figure \ref{s1} shows an idealized schematic diagram of the draping of a purely radial ambient IMF about a fast moving MC and the MC's cross-section in a plane perpendicular to the ecliptic plane. The imbalance in the MC's accumulated azimuthal flux arises due to the reconnection at the MC's front.

An approximate understanding of the erosion of the magnetic flux of MCs due to reconnection with the IMF can be obtained with recourse to the Sweet-Parker model \cite{1957JGR....62..509P,sweet_1958}. In the diffusion region (a small-scale region surrounding the reconnection site) \citeA{2007PhPl...14j2114C} showed that the reconnection rate scales as $E_{rec}\sim V_{1}B_{1}\sim V_{2}B_{2}$, where $V$ and $B$ are the plasma flow speed and magnetic field intensity, respectively, and the subscripts `1' and `2' refer to either side of the inflow region. For steady state reconnection $E_{rec}$ is uniform at the reconnection region, where $E_{rec}$ defines the rate of magnetic flux that is transferred from the inflow to the dissipation region \cite{2018JGRA..123.9150N}. This transferred magnetic flux determines the amount of flux erosion ($F_{er}$) and one expects that $F_{er}$ would scale linearly with the background IMF field strength \cite{2007PhPl...14j2114C}. 

The local Alfv\'{e}n speed ($V_A$) in the plasma near the reconnection site governs the reconnection rate \cite{1973ApJ...180..247P, 2001JGR...106.3715B,2007PhPl...14j2114C}. As $V_A$ decreases with the increasing heliospheric distance, the reconnection rate is expected to be higher closer to the Sun and in the inner heliosphere  (e.g. \citeA{2008JGRA..113.0B08L}). \citeA{2007SoPh..244..115D} found an extended coherent magnetic region just behind an MC and interpreted that this region is formed due to the reconnection process which started $\approx 54$ hrs before the MC was observed. \citeA{2014JGRA..119...26L} suggested that about 47\% to 67\% of the MC erosion is expected to occur within the orbit of Mercury ($\approx 0.39$ AU from the Sun). \citeA{2015JGRA..120...43R} found no correlation between the eroded amount of magnetic flux and the ambient solar wind properties observed in situ at 1 AU. The lack of correlation suggests that most of the reconnection is expected to occur during the MC's passage through the inner heliosphere and not at 1 AU. \par

Large scale open magnetic field originating from the solar corona extends out into the heliosphere and shapes the IMF. Utilizing the observed solar photospheric magnetic field information as boundary condition the heliospheric magnetic field can be reconstructed following the simplest assumption that the solar corona is current-free \cite{Schatten1969}. This approach is known as Potential Field Source Surface extrapolation (PFSS). In the outer radial boundary of the PFSS model the field lines are constrained to be radial which is consistent with the observations of the coronal hole foot-points \cite{Wang1996}. The open magnetic flux ($\phi_{open}$) derived using the Sun's open magnetic field lines can be used to estimate the radial IMF intensity at any heliocentric distance from the Sun as the $Ulysses$ spacecraft measurements show a latitudinal and longitudinal uniformity of the IMF's radial component \cite{2008GeoRL..3522103S,1995Sci...268.1007B,2014ApJ...781...50E}. Therefore, the radial magnetic field intensity ($|B_r|$) can be expressed as $|B_r|=|\phi_{open}|/4\pi r^2$ at a heliocentric distance $r$ ($r>> R_{\odot}$). If $\theta_{HG}$ and $\phi_{HG}$ are the heliographic latitude and longitude, respectively, then the open magnetic flux can be derived by integrating $|B_{r}(R_{ss},\theta_{HG},\phi_{HG})|$ over the surface of a sphere concentric to the Sun with radius $R_{ss}$, known as source surface. Using $\phi_{open}$, several studies have estimated $B_r$ at $r=1$ AU and compared it with the in situ near-Earth radial IMF intensity. The studies found a good correspondence between the estimated and observed values \cite{1988JGR....9311227W,1995ApJ...447L.143W,wang2000long,yeates2010,wang2015coronal}.

The present study aims to establish a connection between the MC's eroded magnetic flux and the IMF intensity. For achieving this we explore the relationship between annual averages of MC's eroded flux and the $\phi_{open}$ over two solar cycles. Also, we present a statistical analysis to investigate the geoeffectiveness of eroded MCs. We describe the data and their sources in section 2 and present our analysis methodology in section 3. Results are presented in section 4 followed by discussion and conclusions in section 5 and section 6, respectively.
\section{Instrumentation and Overview of Data}
We use 64s average level 2 (verified) data of merged interplanetary magnetic field and solar wind parameters (\url{http://www.srl.caltech.edu/ACE/ASC/level2/lvl2DATA_MAG-SWEPAM.html}) from the Advanced Composition Explorer (ACE) \cite{1998SSRv...86....1S} spacecraft's Solar wind Electron, Proton and Alpha Monitor (SWEPAM) \cite{1998SSRv...86..563M} and Magnetic Field Experiment (MAG) \cite{1998SSRv...86..613S} instruments. The selected data is in Geocentric Solar Ecliptic (GSE) coordinate system and available since the year 1998. In GSE coordinate system $\hat{x}_{GSE}$ points toward the Sun, $\hat{z}_{GSE}$ points towards the north pole (of the ecliptic) and $\hat{y}_{GSE}$ completes the right-handed system. To obtain the geomagnetic response of MC we use the geomagnetic activity index ($Dst$) available at the World Data Center, Kyoto (\url{http://wdc.kugi.kyoto-u.ac.jp/dstdir/index.html}) in a temporal resolution of 1 hour.  \par
We obtain the radial magnetic field data from synoptic magnetograms (\url{http://hmi.stanford.edu/data/synoptic.html}) collected by the Michelson Doppler Imager (MDI) \cite{1995SoPh..162..129S} onboard the Solar and Heliospheric Observatory (SOHO) \cite{1995SoPh..162....1D} and the Helioseismic and Magnetic Imager (HMI) \cite{2012SoPh..275..207S} onboard the Solar Dynamics Observatory (SDO) \cite{2012SoPh..275....3P}. The data from MDI and HMI is collected for a period of 1998-April 2010 and May 2010-2018, respectively. We utilize the radial magnetic field data to extrapolate the coronal magnetic field till the source surface at $r = R_{ss}= 2.5R_{\odot}$ using a global PFSS extrapolation model.

\subsection{MC events and criteria for selection}

In general, MCs are identified following the criteria defined by \citeA{1981JGR....86.8893B} which are (1) throughout an MC interval the magnetic field intensity enhances (compared to the ambient solar wind), (2) smooth rotation in both $B_{y,GSE}$ and $B_{z,GSE}$ components exists, (3) the proton temperature falls below the expected solar wind temperature \cite{lopez1986solar}, and (4) the plasma $\beta$ parameter is less than 1. For solar cycle 23 (1998-2007) we consider 37 MCs from a list of 109 MCs selected by \citeA{2015JGRA..120...43R}. The study categorized the MCs as Q1 (both the front and rear boundaries of MCs are well determined), Q2 (any one of the two boundaries is identified without ambiguity) and Q3 (both the boundaries are difficult to identify). Of them, we select only Q1 and Q2 type MCs for our study as the fitting of flux ropes to the MCs is impacted by incorrect boundary selection. The events during 2008-2014 are obtained from the MC list published by \citeA{2015JGRA..120..9221G} and the events of 2015-2018 are selected from the \citeA{Richardson2010} ICME catalogue (\url{http://www.srl.caltech.edu/ACE/ASC/DATA/level3/icmetable2.htm}) after manually identifying each of their boundaries. Thus, we select a total number of 35 events during solar cycle 24. Note that, each of the events selected here is consistent with \citeA{1981JGR....86.8893B}'s definition of MCs. For none of the events, the spacecraft crosses the MC flux ropes parallel to the legs. This is because those cases lead an azimuthal flux asymmetry which may be misinterpreted as a result of erosion \cite{2015JGRA..120...43R}. Also, there exist uncertainties in fitting parameters [e.g. \citeA{2007AnGeo..25.2453M,2003JGRA..108.1356L}] while those MCs are fit with the cylindrical flux rope model. \citeA{2013A&A...556A..50J} introduced a location angle ($\lambda$) measured from the plane ($\hat{y}_{GSE}$,$\hat{z}_{GSE}$) towards the MC axis. It is a proxy for the spacecraft crossing distance from the MC nose. The $\lambda$ evolves monotonically along the MC flux rope with $\lambda=-90^\circ$ in one leg, $\lambda=0^\circ$ at its nose, and $\lambda=90^\circ$ in the other leg. \citeA{2015JGRA..120...43R} estimated that errors in the azimuthal flux imbalance is less than 10\% if $\lambda$ holds a value between $\pm45^\circ$. Thus we study only those MCs which cross the spacecraft trajectory sufficiently close to their apex by selecting $\lambda<|45^\circ|$.    
\par 

\section{Analysis methodology}

Several case studies \cite{2006A&A...455..349D, 2007SoPh..244..115D} show the presence of unbalanced azimuthal magnetic flux in MCs although MCs maintain a classical characteristic. Since the outer part of an MC is mainly affected by reconnection and the azimuthal flux is much greater than the axial flux in there \cite{2015JGRA..120...43R}, the asymmetry is mainly observed in the MC's azimuthal magnetic flux. To estimate this asymmetry we employ the ``direct method" developed by \citeA{2006A&A...455..349D}. 

\subsection{Measurement of MC's eroded magnetic flux}

The direct method uses the principle of magnetic flux conservation ($\nabla \cdot \mathbf{B}=0$) across the plane formed by the spacecraft trajectory and MC axis to determine and analyse the MC's accumulated azimuthal magnetic flux. The MC axis orientation is defined by a latitude angle $\theta$ (the angle between MC's axis and the ecliptic plane) varying from $-90^\circ$ to $90^\circ$ and a longitude angle $\phi$ (the angle between the projection of MC's axis on the ecliptic plane and the Earth-Sun direction) varying from $0^\circ$ to $360^\circ$. Using $\theta$ and $\phi$, we rotate the observed MC from GSE to the local MC coordinate defined by \citeA{2006A&A...455..349D} and derive $\mathbf{B}_{cloud}$ and $\mathbf{V}_{cloud}$ from $\mathbf{B}_{GSE}$ and $\mathbf{V}_{GSE}$. To obtain $\theta$ and $\phi$ we use a constant-$\alpha$ force-free flux rope model \cite{1988JGR....93.7217B, 1986AdSpR...6..335M} to least-square fit the observed MCs. The model follows the equation $\nabla \times \mathbf{B}=\alpha \mathbf{B}$, where $\alpha$ is a constant and allows self-similar expansion of an MC. The Lundquist's solution to the equation \cite{1990JGR....9511957L} provides the modeled magnetic field vectors in the cylindrical coordinate system. After performing a series of iterations through proper adjustment of model parameters based on minimizing the difference between modeled and observed magnetic field vectors, a final set of best-fit parameters including $\theta$ and $\phi$ are ascertained (e.g, \citeA{2007AnGeo..25.2453M, 2012JGRA..117.1101M}).

\begin{figure}[htbp]
 \centering
 \includegraphics[width=0.7\textwidth]{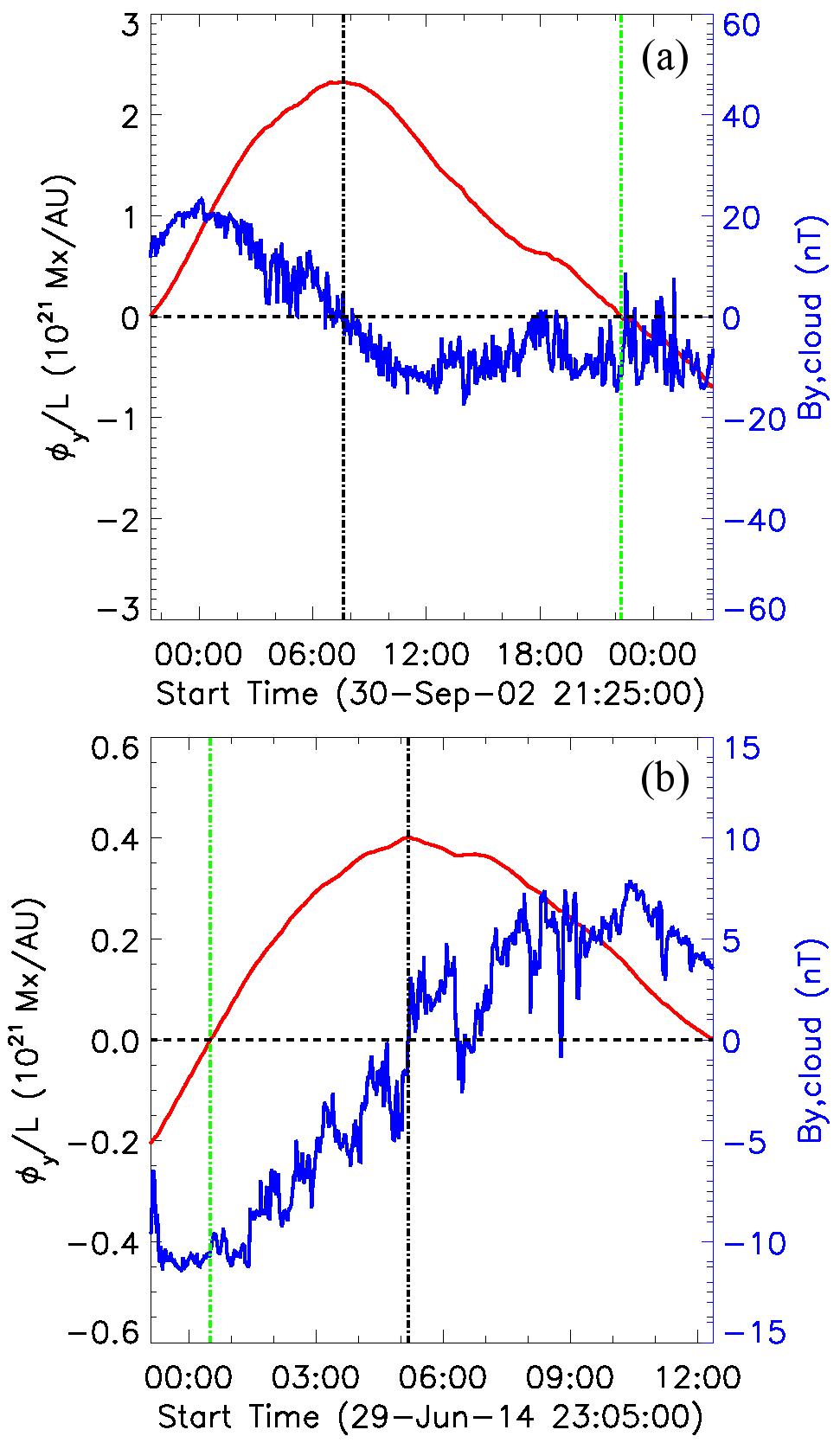}
 \caption{Plots depicting $B_{y,cloud}$ (blue) and $\phi_{y}/L$ (red) variations of two MCs during their passage past the spacecraft. The black and green vertical dashed-dotted lines represent $t_{center}$ and the zero crossing of $\phi_{y}/L$ curves, respectively. (a) For an event where the flux erosion is at the MC-front. Here $\phi_{az} = 3.02\times10^{21}$ Mx/AU and eroded flux = $6.98\times10^{20}$ Mx/AU. (b) For an event where the flux erosion is at MC-rear. Here $\phi_{az} = 6.06\times10^{20}$ Mx/AU and eroded flux =  $2.04\times10^{20}$ Mx/AU.}
 \label{s2}
\end{figure}

If $\theta$, $\phi$ and the time of MC front ($t_{front}$) and rear boundaries ($t_{rear}$) are known, the accumulated azimuthal magnetic flux per unit length ($\frac{\phi_{y}(t_1,t_2)}{L}$) can be measured using
\begin{equation}
\label{eqn1}
    \frac{\phi_{y}(t_1,t_2)}{L}=\int_{t_1}^{t_2}B_{y,cloud}(t)V_{x,cloud}dt,
\end{equation}
where $L$ is the total length of the MC axis till 1 AU, $B_{y,cloud}$ and $V_{x,cloud}$ are the MC's magnetic field and velocity components computed in the local cloud coordinate system, respectively. The times $t_1$ and $t_2$ represent generic times defining the boundaries of the integral. The normalized perpendicular distance between spacecraft trajectory and MC axis is called the impact parameter ($Y_0$). The accumulated azimuthal flux is more precisely estimated when $Y_0$ is low. In Figure \ref{s2} we plot the time evolution of $B_{y,cloud}$ (blue curve) along with $\phi_{y}/L$ (red curve) of two MCs during their passage through the spacecraft. Figure \ref{s2}a shows an MC that is eroded at the front edge. The parameter $\phi_{y}/L$ is calculated considering $t_1=t_{front}$ and $t_2=t_{rear}$. Due to erosion its accumulated azimuthal flux shows an excess at its rear. The total azimuthal flux per unit length ($\phi_{az}$) that was originally present in that MC (before erosion) is estimated by the unsigned value of $\phi_{y}(t_{center},t_{rear})/L$. Here $t_{center}$ (shown in dashed-dotted black vertical lines in Figures \ref{s2}a and \ref{s2}b) is the time when $B_{y,cloud}$ changes its sign and results in a maximum value of unsigned $\phi_{y}/L$. The time $t_{center}$ is the time when closest approach of the spacecraft to the MC axis is achieved. The eroded azimuthal flux (per unit length) of the MC is represented by the absolute value of $\phi_{y}/L$ at $t_{rear}$. The second MC (Figure \ref{s2}b) is eroded at its rear and shows a flux imbalance at its front. Here $\phi_{y}/L$ is calculated considering $t_1=t_{rear}$ and $t_2=t_{front}$. For this case $\phi_{az}$ is estimated by the unsigned value of $\phi_{y}(t_{center},t_{front})/L$. The eroded flux of the MC is defined by the absolute value of $\phi_{y}/L$ at $t_{front}$. We next define $\phi_{erod}$ as the eroded flux normalized to $\phi_{az}$. The dashed-dotted green vertical lines show the estimated rear (Figure \ref{s2}a) and front (Figure \ref{s2}b) boundaries of the remaining flux-rope when it is observed in situ.

\subsection{Estimation of the Sun's open flux} 

An estimate of the Sun's open flux is traditionally made based on potential field extrapolation of the Sun's coronal field (satisfying $\nabla \times \mathbf{B} = 0$, \citeA{Schatten1969}). Following \citeA{van_Ballegooijen_2000} and using a module developed by \citeA{anthony_yeates_2018_1472183} we extrapolate the coronal magnetic field utilizing solar surface synoptic maps from MDI and HMI as the lower boundary condition. The unsigned open flux ($|\phi_{open}|$) is estimated at $R_{ss}=2.5R_{\odot}$. We obtain a strong positive correlation (with linear correlation coefficient, $r_p$= 0.89 at 99.99\% confidence level) between the yearly averaged IMF intensity observed in situ at 1 AU and $|\phi_{open}|$. Therefore, $|\phi_{open}|$ can be used as a proxy of the IMF intensity. 

 \section{Analysis and Results} 
\begin{figure}[htbp]
\centering
\includegraphics[height=0.85\textheight]{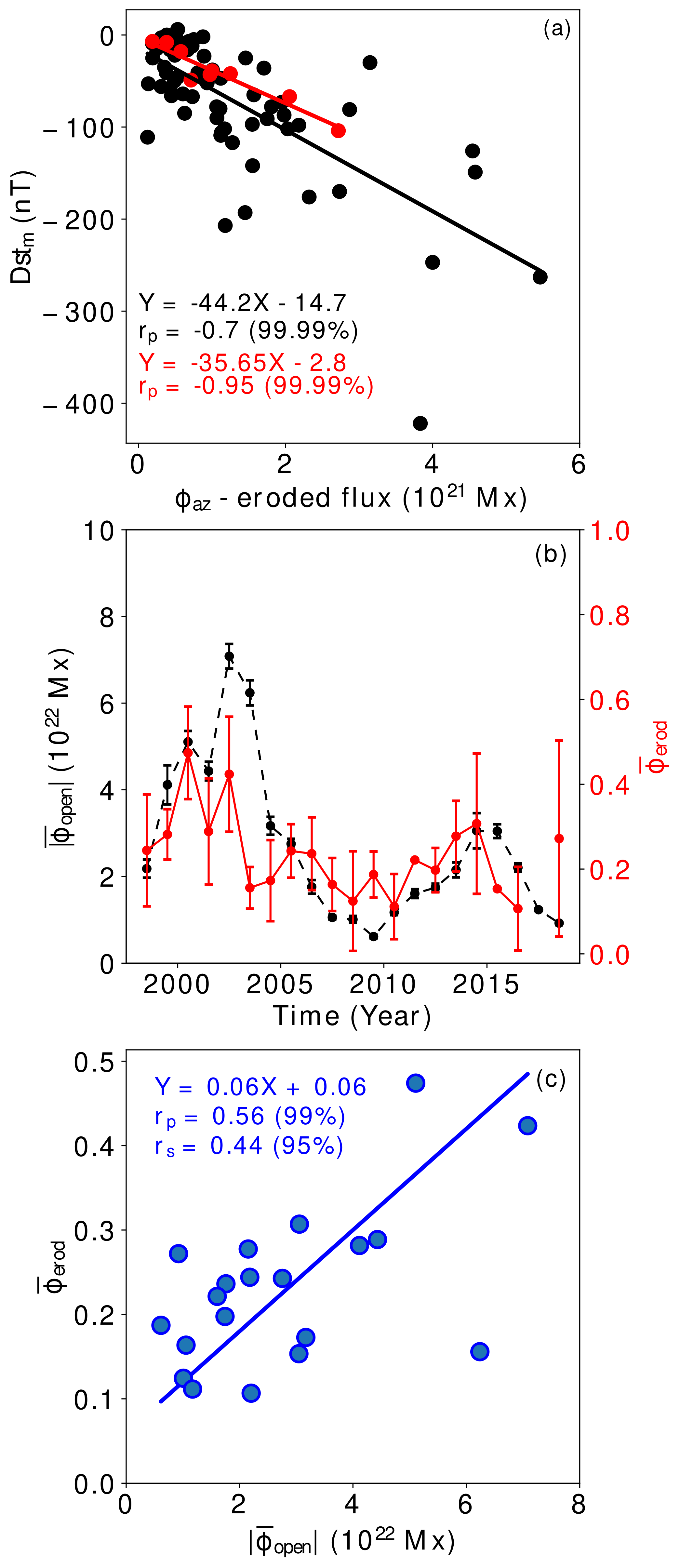}

 \caption{(a) Scatter plot between the remaining flux of MCs and their geoeffectivness as quantified by the minimum $Dst$ index ($Dst_m$). Least-squares fit considering all events (red and black dots) and the correlation are shown in black. For a subset of events (red dots only) as described in the text, the corresponding least-squares fit and correlation are depicted in red. (b) $\bar{\phi}_{erod}$ and $\bar{|\phi}_{open}|$ plotted against each year for solar cycle 23 and 24 in continuous red and dashed black curves, respectively. Vertical lines in corresponding colors represent the error bars for $\bar{\phi}_{erod}$ and $\bar{|\phi}_{open}|$. (c) Scatter plot of $\bar{|\phi}_{open}|$ versus $\bar{\phi}_{erod}$. The over plotted blue solid line shows the least-squares fit to the data points. The correlation is mentioned in blue. }
 \label{s3}
\end{figure}

 By applying the direct method to all the selected MCs we derive $\phi_{az}$, $\phi_{erod}$, and the position of the flux imbalance (front or back) for each of the MCs. In Figure \ref{s3} (a) we demonstrate the effect of MC's erosion on its geoeffectiveness by plotting MC's remaining azimuthal flux ($\phi_{az}-$ eroded flux) versus its geomagnetic response (quantified by $Dst_m$-- the minimum value of $Dst$ during an MC's passage). The correlation between the remaining flux and $Dst_m$ ($r_p= -0.7$ at 99.99\% confidence) is slightly lower than the correlation between total flux and $Dst_m$ ($r_p= -0.76$ at 99.99\% confidence) if all events are considered. However, it is to be noted that the impact of MC's azimuthal flux erosion on its geomagnetic response is expected only if $B_{z,GSE}$ has a negative value at the reconnection site and the MC's azimuthal field component corresponds to $B_{z,GSE}$ (i.e., $\theta, \lambda \sim 0^\circ$) near the Earth (i.e., at 1 AU). Our sample has 9 MCs (depicted by red dots in Figure \ref{s3} (a)) which have negative $B_{Z,GSE}$ at their reconnection sites and $|\theta|, |\lambda| \leq 30^\circ$ (i.e., close to the above conditions, while still providing reasonable sample size). For these events the correlation between the remaining flux of MCs and $Dst_m$ is higher ($r_p = -0.95$  at 99.99\% confidence) compared to that between $\phi_{az}$ and $Dst_m$ ($r_p = -0.88$ at 99.99\% confidence). This reveals the impact of flux erosion on the geoeffectiveness of MCs.
\par      
 
  We notice that out of 72, 40 MCs show the flux imbalance at rear with an average value of $\phi_{erod}$ $\sim 0.26$ and 32 MCs show flux imbalance at front with an average $\phi_{erod}$ $\sim 0.23$. The average eroded flux irrespective of the erosion position during solar cycle 23 and 24 are found to be 0.276 and 0.22, respectively. Our result matches well with the value of normalized average eroded flux found by \citeA{2015JGRA..120...43R} for the events during solar cycle 23. MCs showing flux imbalance at the front or rear boundaries are named as $MC_F$ and $MC_{R}$, respectively.  
 In Table \ref{table1}, we show the number of total MCs (column 2), $MC_{R}$ (column 3), $MC_F$ (column 4), annual averages of $\phi_{erod}$ (column 5), and unsigned $\phi_{open}$ (column 6) with their standard deviations corresponding to each year mentioned in column 1.      \par
 
 \subsection{Comparison between the Sun's unsigned open flux and MC flux erosion}

 The erosion of magnetic flux in an MC depends on the rate of reconnection between the magnetic field of the MC and IMF. As it is stated before that the reconnection rate scales with the IMF intensity, a correlation is expected between $\phi_{erod}$ and the prevailing IMF strength whose proxy is the unsigned $\phi_{open}$. We define the yearly average of the normalized eroded azimuthal flux ($\bar{\phi}_{erod}$) by averaging over $\phi_{erod}$ with the total number of events ($N_{MC}$) in each year. In Figure \ref{s3}b, we plot $\bar{\phi}_{erod}$ and the yearly average of unsigned $\phi_{open}$ ( $\bar{|\phi}_{open}|$) in solid red and dashed black curves, respectively, along with error bars over the period encompassing solar cycle 23 and 24. The error bars are calculated by estimating, $\frac{\sigma}{ \sqrt{N_{event}}}$, where $\sigma$ is the standard deviation and $N_{event}$ represents number of MC events for the eroded flux and number of PFSS extrapolations for the calculation of the open flux in each year. A correlation study is performed between $\bar{\phi}_{erod}$ and $\bar{|\phi}_{open}|$. We obtain a Pearson correlation coefficient ($r_p$) of $0.56$ and a Spearman's rank correlation coefficient ($r_s$) of $0.44$, at 99\% and 95\% confidence levels, respectively. To confirm our result we compute $r_p$ including the parameters' error bar information and find the value of $r_p$ as $0.51 \pm 0.03$ at 98\% confidence level. In Figure \ref{s3}c, we show a scatter plot between $\bar{\phi}_{erod}$ and $\bar{|\phi}_{open}|$. The observed correspondence between the eroded flux and the Sun's open flux indicates an underlying linear relationship between the reconnection rate and the IMF intensity. To relate these two parameters, we perform a least-squares fit to $\bar{|\phi}_{open}|$ versus $\bar{\phi}_{erod}$ which gives the following equation
 \begin{equation}
 \label{eqn7}
     \bar{\phi}_{erod}=0.06\bar{|\phi}_{open}|+0.06,
 \end{equation}
where $\bar{|\phi}_{open}|$ is in units of $10^{22} Mx$. The solid blue line over plotted on the scatter plot represents the least-squares fitting.\par

 \section{Discussion}
 
To establish a correspondence between MC's eroded flux and the IMF intensity, we study the dependency of the MC's eroded flux on the Sun's open flux over two solar cycles. 
We find that about $44\%$ of the total MCs studied here shows an imbalance at their rear boundaries. A numerical simulation of reconnection between an MC and solar wind performed by \citeA{2003JGRA..108.1023S} shows an enhancement in reconnection with the increasing relative speed of the MC compared to the solar wind. If an MC is followed and compressed by a fast solar wind, reconnection may occur at the MC's rear boundary. Such high speed solar wind primarily originates from polar coronal holes and may extend towards low latitudes \cite{1998GeoRL..25.2999F}. Also, MCs can be compressed by co-rotating interaction regions whose formation is related to the presence of coronal holes \cite{2010ApJ...719.1385R}.\par
In our study we obtain the orientation of an MC axis by fitting the observed MC using a force free cylindrical flux rope model. The model is impacted by several factors, such as, incorrect boundary selection \cite{2003JGRA..108.1356L}, non circular MC cross section \cite{2011ApJ...732..117S,2011ApJ...731..109S} and high impact parameter values \cite{2013A&A...550A...3D,2004JASTP..66.1321R}. \citeA{2015JGRA..120...43R} showed that the flux rope fitting method has a tendency of deriving a consistently lower value of eroded flux because the model is based on axisymmetric geometry. Also, a high impact parameter lowers the total azimuthal magnetic flux that causes an overestimation of the normalized eroded flux. \par

The imbalance in MC azimuthal flux accumulated along the spacecraft path may occur due to various other reasons. When a high-speed CME propagates from Sun to Earth it undergoes strong deceleration \cite{2007JGRA..112.8107J}. If a cool, dense filament material is located at the base of the CME, the momentum of the filament results in its forward movement through the decelerating CME. Thus, the protruding filament drives flow leading to a sideways transport of the CME's poloidal field. This can generate an azimuthal flux imbalance in the CME flux rope by the time the CME reaches the Earth. \citeA{2014JGRA..119.5449M} explained such flux imbalances through simulation of a three-part structure CME observed on 20 January 2005. In such cases, the eroded magnetic flux amount does not depend on the rate of reconnection between the MC and the IMF. This may also contribute to a lower correlation between the amount of eroded MC flux and $\phi_{open}$. Nonetheless, we find the correlation to be significant and the underlying relationship between $\phi_{erod}$ and $\phi_{open}$ to be linear.
 
\section{Conclusions}
We establish a relationship between the azimuthal magnetic flux erosion of MCs and the Sun's unsigned open flux that regulates the IMF intensity. Utilizing the direct method we investigate 72 MCs spanning solar cycles 23 and 24 to estimate the magnetic flux imbalance and erosion. We find that on average $28\%$ and $22\%$ of the total azimuthal magnetic flux of MCs of solar cycle 23 and 24, respectively, are eroded during their propagation through the inner heliosphere. The reconnection causing this erosion may occur on either side of the MC boundaries and peel off an almost similar amount of magnetic flux. We compare the annual averages of solar open flux to the average fraction of the eroded magnetic flux of MCs over the past 21 years (from 1998 to 2018) and find a significant positive correlation with an underlying linear relationship. Since the solar open flux is strongly correlated with the IMF intensity, this is suggestive of the latter's role in MC flux erosion. 

We note that the solar open flux is governed by the emergence and redistribution of active region magnetic fields on the Sun's surface due to surface flux transport processes -- a crucial component of the solar dynamo mechanism \cite{2018NatCo...9.5209B}. Given that the solar wind dispersed open flux determines the ambient heliospheric magnetic field (IMF), the latter's solar cycle modulation provides a novel pathway via which the large-scale solar cycle can govern flux erosion and thus possibly the geoeffectiveness of interplanetary magnetic clouds.  
 
\acknowledgments
The Center of Excellence in Space Sciences India (CESSI) is funded by the Ministry of Human Resource Development, Government of India under the Frontier Areas of Science and Technology scheme. S.D. acknowledges funding from the DST-INSPIRE program of the Government of India. The authors are grateful to Benoit Lavraud for insightful discussions and suggestions. We thank Katsuhide Marubashi for making available the constant-$\alpha$ force free cylindrical flux rope model. We thank an anonymous referee for useful comments. The authors acknowledge the use of data from the SDO/HMI, SOHO/MDI, ACE instruments, and the World Data Center for Geomagnetism, Kyoto. 


\begin{sidewaystable}
 \caption{Total number of MCs, $MC_{R}$, $MC_F$ in each year, annual average of normalised eroded flux ($\bar{\phi}_{erod}$), and unsigned  $\phi_{open}$ ($\bar{\phi}_{open}$) along with their standard deviations over the two solar cycles (1998-2018). }
 \centering
 \begin{tabular}{c c c c c c }
 \hline
Year&$N_{MC}$&$MC_{R}$&$MC_F$&$\bar{\phi}_{erod}$& $\bar{\phi}_{open}$  \\
&$\#$&$\#$&$\#$&&$(\times10^{22}$Mx)\\
 \hline
      1998&       5&       2&       3&0.24$\pm$0.13&2.18$\pm$0.21\\
    1999&       3&       2&       1&0.28$\pm$0.06&4.12$\pm$0.45\\
    2000&       5&       4&       1&0.47$\pm$0.11&5.11$\pm$0.25\\
    2001&       4&       2&       2&0.29$\pm$0.13&4.43$\pm$0.22\\
    2002&       3&       3&       0&0.42$\pm$0.14&7.08$\pm$0.28\\
    2003&       2&       0&       2&0.16$\pm$0.05&6.24$\pm$0.29\\
    2004&       5&       4&       1&0.17$\pm$0.10&3.17$\pm$0.21\\
    2005&       3&       1&       2&0.24$\pm$0.06&2.76$\pm$0.11\\
    2006&       3&       3&       0&0.24$\pm$0.09&1.76$\pm$0.16\\
    2007&       2&       0&       2&0.16$\pm$0.06&1.06$\pm$0.05\\
    2008&       2&       2&       0&0.12$\pm$0.12&1.01$\pm$0.08\\
    2009&       4&       1&       3&0.19$\pm$0.05&0.61$\pm$0.04\\
    2010&       3&       1&       2&0.11$\pm$0.08&1.17$\pm$0.06\\
    2011&       1&       1&       0&0.22$\pm$0.00&1.60$\pm$0.10\\
    2012&       9&       4&       5&0.20$\pm$0.05&1.74$\pm$0.09\\
    2013&      10&       6&       4&0.28$\pm$0.08&2.15$\pm$0.17\\
    2014&       3&       0&       3&0.31$\pm$0.17&3.06$\pm$0.41\\
    2015&       1&       1&       0&0.15$\pm$0.00&3.05$\pm$0.16\\
    2016&       2&       2&       0&0.11$\pm$0.10&2.20$\pm$0.10\\
    2017$^a$&       0&       0&       0&--&1.23$\pm$0.03\\
    2018&       2&       1&       1&0.27$\pm$0.23&0.92$\pm$0.05\\
 \hline
\multicolumn{2}{l}{$^{a}$None of the observed MCs follow our selection criteria.}
\label{table1}
 \end{tabular}
\end{sidewaystable}

\end{document}